# Can Machine Learning Create an Advocate for Foster Youth?

Meredith Brindley
Think of Us
Richmond, VA, USA
meredith@thinkof-us.org

James Heyes
Think of Us
Richmond, VA, USA
james.heyes@thinkof-us.org

Darrell Booker
Think of Us
Richmond, VA, USA
darrell@thinkof-us.org

## ABSTRACT

Statistics are bleak for youth aging out of the United States foster care system. They are often left with few resources, are likely to experience homelessness, and are at increased risk of incarceration and exploitation. The Think of Us platform is a service for foster youth and their advocates to create personalized goals and access curated content specific to aging out of the foster care system. In this paper, we propose the use of a machine learning algorithm within the Think of Us platform to better serve youth transitioning to life outside of foster care. The algorithm collects and collates publicly available figures and data to inform caseworkers and other mentors chosen by the youth on how to best assist foster youth. It can then provide valuable resources for the youth and their advocates targeted directly towards their specific needs. Finally, we examine machine learning as a support system and aid for caseworkers to buttress and protect vulnerable young adults during their transition to adulthood.

## 1.INTRODUCTION

The foster care system is the last lifeline for the nation's most vulnerable young people. Most children or teens entering foster care do so due to neglect, physical abuse, parental substance abuse, or a care taker's inability to provide for them. Currently, there are about 428,000 children in foster care across the United States [8]. Each year, around 25,000 to 30,000 of these young people will age out of the system without permanent families [8]. Upon exit from the foster care system without a family or support network, youth face a higher risk of homelessness, exploitation, teen pregnancy, and incarceration. The American Journal of Public Health indicates that 31-46% of youth aging out of the foster system will become homeless by the age of 26 [3]. In a University of Chicago study from 2011, more than 40% of former foster males tracked in Iowa, Illinois, and Wisconsin, reported being incarcerated, compared to 10% in the population of young adults at large in those states [2]. Overall, foster youth were nearly six times more likely to report being convicted of a crime as an adult while in foster care than the rest of their cohort [2]. Females in foster care who had been arrested were also more likely to have been pregnant than those foster youths who had never been arrested [6]. In general, by age 19, regardless of history of maltreatment, religious faith, or academic performance, 55% of females in foster care had been pregnant at some point [6].

Foster youth without a permanent home are also exceptionally vulnerable to sex trafficking. The National Center for Missing and Exploited Children estimates that six out of ten children involved in child sex trafficking have been in foster care [5].

Most states use a joint program called the Statewide Automated Child Welfare Information System (SACWIS) and the Tribal Automated Child Welfare Information System (TACWIS) to organize caseworker caseloads [7]. Though intended for tracking data on foster youth to better assist them, it has caused late payments to foster families, and can cause complications for states during self-audits, unnecessarily increasing the need for federal oversight. Many caseworkers say they find themselves spending more time at the computer and less time assisting and protecting youth under their care [4]. Machine learning has the potential to streamline the process of navigating these systems for caseworkers, and to help youth better advocate for themselves and their needs while transitioning out of the foster care system.

## 2.THE THINK OF US PLATFORM

The Think of Us Platform enables youth to create goals for their life and add adults to support their goals (Figure 1). It is being piloted in Santa Clara County, California and in the state of Nebraska.

The platform features curated content for youth in areas including financial management, family planning, physical and mental health (Figure 2). The goals youth set allow caseworkers to create a Transition Independent Living Plan (TILP) required by the state. [9] The platform allows youth to continuously edit their TILP plan without the need for an appointment with the caseworker, attorney or deputy probation officer. Without the Platform, a youth would have to go to one of those figures to update their TILP, who would then input this data in SACWIS and TACWIS at a later time, which can be cumbersome.

The Supporters function allows youth to select supportive adults to assist with these goals. These youth-identified supportive adults may be school teachers, coaches or other people who the youth feels can provide support with a specific goal. Supporters are assigned on individual goals rather than in general to allow the youth to consider supporters on a case-by-case basis rather than a general support which they may not feel they have.

We are introducing machine learning into the Platform to support youth with their TILPs, as a way to lighten the load of caseworkers, and to enhance privacy protections for the young people, while simultaneously assisting them with the transition into adulthood.

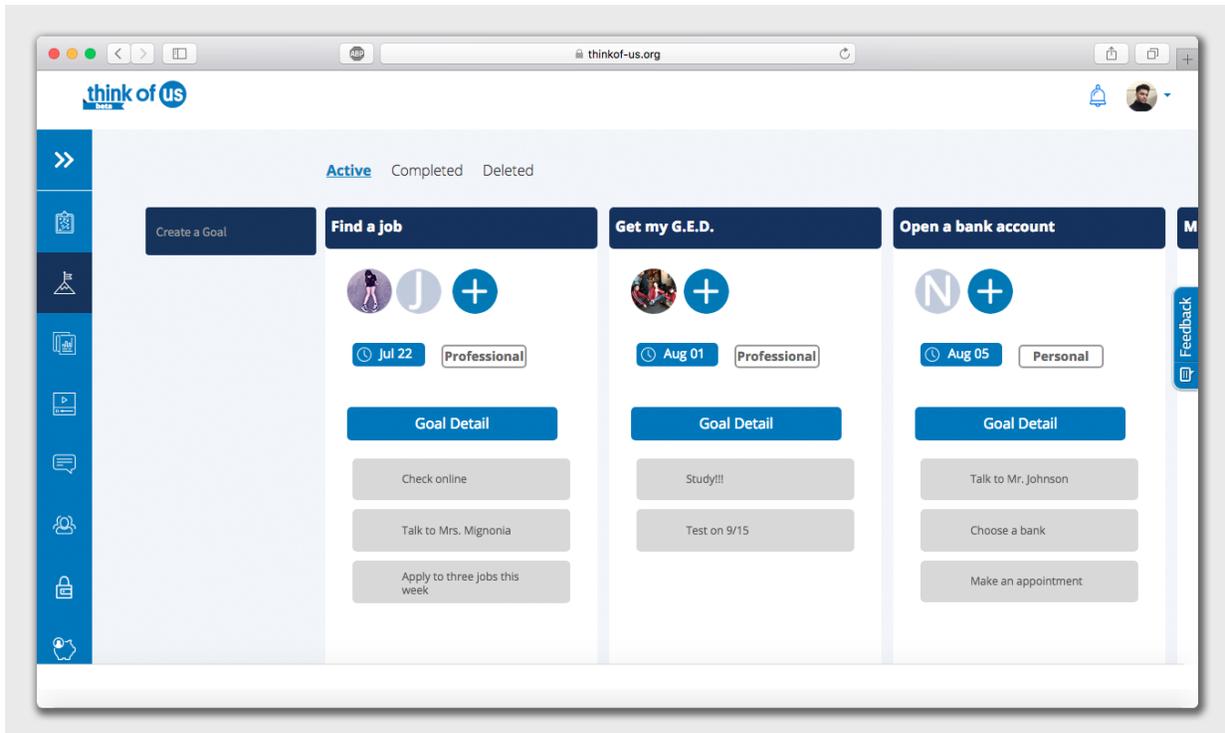

**Figure 1:** Youth can create goals and add supporters to their goals as they transition to adulthood.

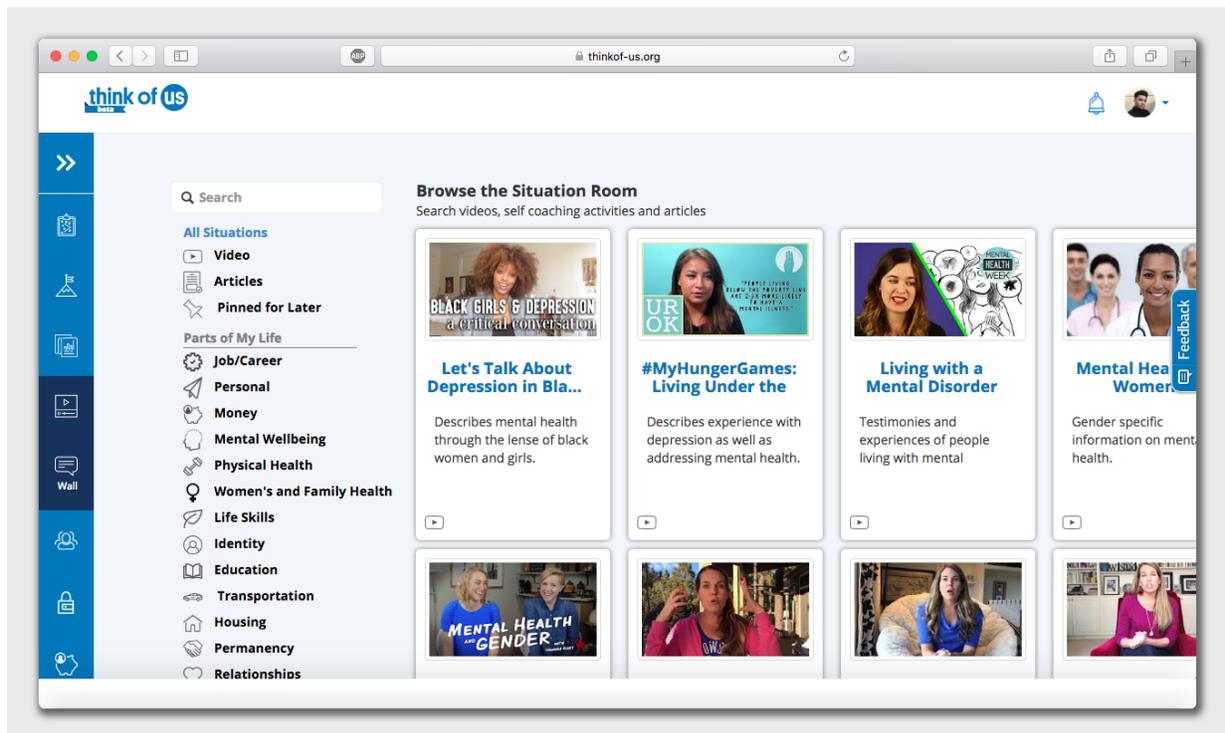

**Figure 2**: Youth can access curated content that in a variety of topics relevant to aging out of the foster care system.

## 3. MACHINE LEARNING

The recent addition of machine learning to the platform allows the system to reach out to youth with recommendations for goals. These include goals that meet immediate needs – e.g. getting a job, finding adequate housing, or advancing their education. The recommendations include action steps that break down the suggested goal into more manageable pieces.

If a youth is in a crisis, s/he can communicate with the chatbot function which by pulling housing data, location, price indices for the area and reasonable budget goals (if the youth has not provided budget information (Figure 3)), can recommend a living situation that meets the present needs of the youth. After parsing, through the youth's inputs with the NLP library Spacy, Keras and Google's Tensorflow, suggestions might include information on temporary shelters, apartments, and suggestions for how to apply for affordable housing in the specific geographic locations of interest to the youth.

The youth's supporters now receive emails and system notifications with suggestions and information on how to assist them in accomplishing a goal if they have been added as a supporter to that goal; the system provides the supporter access to the information necessary to properly support the youth.

As with the supporters' side of the platform, the platform can email the caseworker with an update or news on the youth's progress with goals and suggest ways the caseworker might assist. It can also determine if a youth within their caseload is not receiving sufficient attention and support, and prompt the caseworker to look into the situation. Additionally, if the system identifies language that indicates that the youth appears to be in need of emergency assistance but has not used the request feature, the platform can prompt the caseworker to examine their situation and suggest a course of action appropriate to that specific youth.

## 4. CONCLUSION

A foster youth in crisis needs support and a clear set of steps to help them achieve stability and safety. Caseworkers are often overtaxed, especially by paperwork and navigating tedious information systems. Many foster youths feel they lack support from their community and are in search of ways to become better advocates for themselves. We propose the Think of Us platform and its machine learning capabilities as a way for foster youth to effectively establish goals, consider who could provide support on a specific goal, build a support network, and make efficient contact with their caseworker. With the data we have collected from current youth and alumni of the foster care system, we are employing machine learning to aid foster youth in recommending goals, contacting supporters and caseworkers, and getting the youth into a stable living situation, especially during crisis. Machine learning allows the application to grow and evolve to better serve the needs of foster youth aging out and preparing to age out of the foster care system. The Think of Us application is now being piloted in the foster care systems in the state of Nebraska and Santa Clara County, California, and feedback from real-life implementation will be used to further improve the application so that foster youth and their advocates can create clear, effective plans when transitioning into adulthood.

## 5. ACKNOWLEDGEMENTS

We thank the foster youth, foster alumni and caseworkers who have offered their stories and contributed to the beginning of Think of Us Platform database. We also thank Sixto Cancel for his vision and support in the addition of machine learning to the Think of Us Platform, and Daniel Torraca his design work on the Platform.

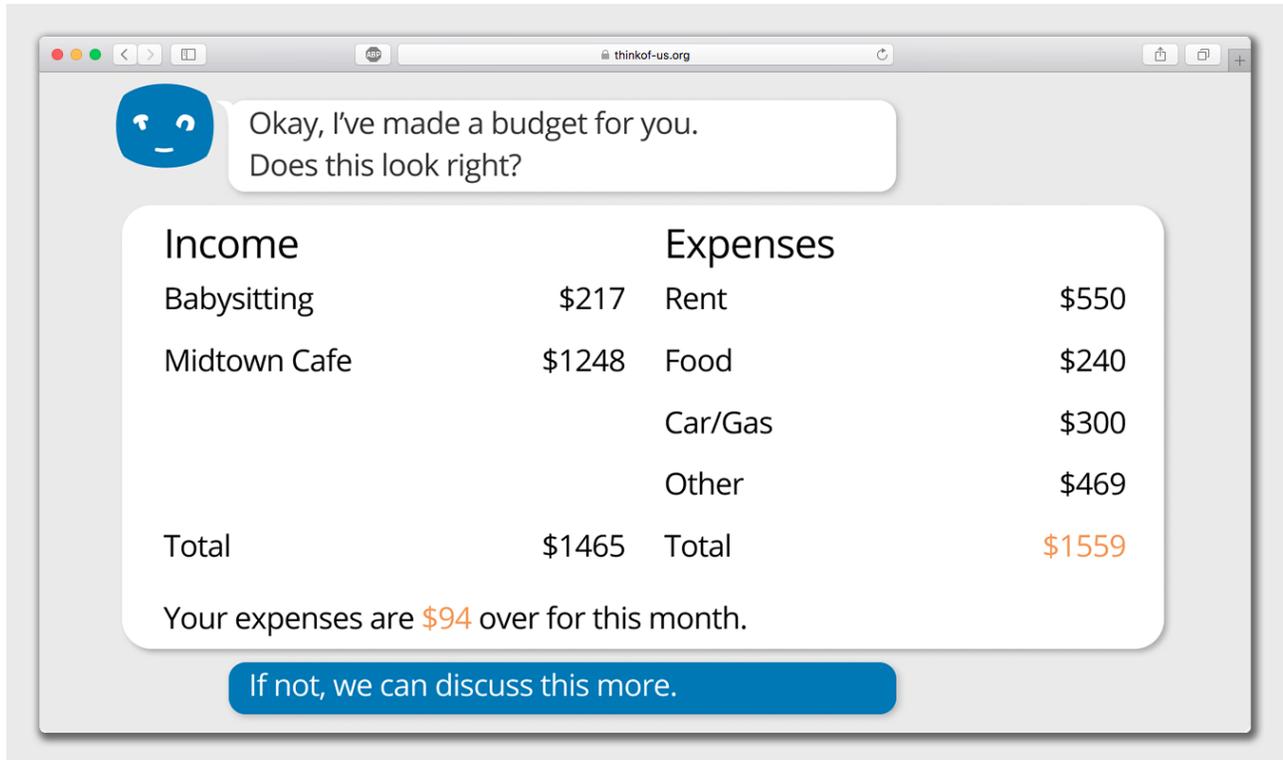

**Figure 3:** Youth can interact with a chatbot that suggest budgets based on area specific data and the user input data.